\documentclass[12pt,a4paper]{article}
\pdfoutput=1

\usepackage{ifthen} \newboolean{pdflatex}
\setboolean{pdflatex}{true} 

\newboolean{articletitles}
\setboolean{articletitles}{true} 

\newboolean{uprightparticles}
\setboolean{uprightparticles}{false} 

\newboolean{inbibliography}
\setboolean{inbibliography}{false} 

\newboolean{paperconf}
\setboolean{paperconf}{false} 

\ifthenelse{\boolean{paperconf}}{\def\paperauthors{LHCb collaboration}
}{\def\paperauthors{R.~Aaij, S.~Benson, M.~De~Cian, A.~Dziurda, C.~Fitzpatrick, E.~Govorkova, O.~Lupton, R.~Matev, S.~Neubert, A.~Pearce, H.~Schreiner, S.~Stahl, M.~Vesterinen}
}
\def\paperasciititle{A comprehensive real-time analysis model at the LHCb experiment}
\def\papertitle{\paperasciititle} \def\paperkeywords{{High Energy Physics}, {LHCb}, {Trigger}, {Real-time analysis}} \def\papercopyright{CERN on behalf of the LHCb collaboration}
\def\paperlicence{CC-BY-4.0}
\def\paperlicenceurl{https://creativecommons.org/licenses/by/4.0/}

\usepackage[top=1in, bottom=1.25in, left=1in, right=1in]{geometry}

\usepackage{microtype}
\usepackage{lineno}  \usepackage{xspace} \usepackage{caption} 

\usepackage{authblk}

\usepackage{graphicx}  \usepackage{color}
\usepackage{colortbl}
\usepackage{booktabs}
\graphicspath{{./figures/}} \usepackage{tikz}
\usetikzlibrary{shapes, arrows, positioning, calc, fit}

\definecolor{bsblue}{HTML}{D1ECF1}
\definecolor{bsblueoutline}{HTML}{BEE5EB}
\definecolor{bsgreen}{HTML}{D4EDDA}
\definecolor{bsgreenoutline}{HTML}{C3E6CB}
\definecolor{bsgrey}{HTML}{EEEEEE}
\definecolor{bsgreyoutline}{HTML}{CCCCCC}

\usepackage{amsmath} \usepackage{amssymb}
\usepackage{amsfonts}
\usepackage{upgreek} 

\newcommand*\patchAmsMathEnvironmentForLineno[1]{\expandafter\let\csname old#1\expandafter\endcsname\csname #1\endcsname
\expandafter\let\csname oldend#1\expandafter\endcsname\csname
end#1\endcsname
 \renewenvironment{#1}   {\linenomath\csname old#1\endcsname}   {\csname oldend#1\endcsname\endlinenomath}}
\newcommand*\patchBothAmsMathEnvironmentsForLineno[1]{  \patchAmsMathEnvironmentForLineno{#1}  \patchAmsMathEnvironmentForLineno{#1*}}
\AtBeginDocument{\patchBothAmsMathEnvironmentsForLineno{equation}\patchBothAmsMathEnvironmentsForLineno{align}\patchBothAmsMathEnvironmentsForLineno{flalign}\patchBothAmsMathEnvironmentsForLineno{alignat}\patchBothAmsMathEnvironmentsForLineno{gather}\patchBothAmsMathEnvironmentsForLineno{multline}\patchBothAmsMathEnvironmentsForLineno{eqnarray}}

\usepackage[binary-units=true]{siunitx}
\usepackage[nolist,nohyperlinks]{acronym}

\usepackage{hyperxmp}

\usepackage[pdftex]{hyperref}

\usepackage[all]{hypcap}

\usepackage{xspace} 
\usepackage{upgreek}

\def\lhcb {\mbox{LHCb}\xspace}
\def\atlas  {\mbox{ATLAS}\xspace}
\def\cms    {\mbox{CMS}\xspace}

\def\runone {\mbox{Run 1}\xspace}
\def\runtwo {\mbox{Run 2}\xspace}
\def\runthree {\mbox{Run 3}\xspace}

\def\velo   {VELO\xspace}
\def\rich   {RICH\xspace}
\def\richone {RICH1\xspace}
\def\richtwo {RICH2\xspace}
\def\ttracker {TT\xspace}
\def\intr   {IT\xspace}

\def\ot     {OT\xspace}

\def\spd    {SPD\xspace}
\def\presh  {PS\xspace}
\def\ecal   {ECAL\xspace}
\def\hcal   {HCAL\xspace}
\def\MagUp {\mbox{\em Mag\kern -0.05em Up}\xspace}

\def\lzero  {L0\xspace}
\def\hlt    {HLT\xspace}
\def\hltone {HLT1\xspace}
\def\hlttwo {HLT2\xspace}

\ifthenelse{\boolean{uprightparticles}}{

 \def\Ppi         {\ensuremath{\uppi}\xspace}

 \def\Ppsi        {\ensuremath{\uppsi}\xspace}

 \def\PDelta      {\ensuremath{\Delta}\xspace}                 
 \def\PXi      {\ensuremath{\Xi}\xspace}                 
 \def\PLambda      {\ensuremath{\Lambda}\xspace}                 
 \def\PSigma      {\ensuremath{\Sigma}\xspace}                 
 \def\POmega      {\ensuremath{\Omega}\xspace}                 
 \def\PUpsilon      {\ensuremath{\Upsilon}\xspace}

 \def\PB      {\ensuremath{\mathrm{B}}\xspace}                 
                  
 \def\PD      {\ensuremath{\mathrm{D}}\xspace}

 \def\PJ      {\ensuremath{\mathrm{J}}\xspace}                 
 \def\PK      {\ensuremath{\mathrm{K}}\xspace}

 \def\Pc      {\ensuremath{\mathrm{c}}\xspace}

 \def\Pi      {\ensuremath{\mathrm{i}}\xspace}

 \def\Pp      {\ensuremath{\mathrm{p}}\xspace}

}
{

 \def\Ppi         {\ensuremath{\pi}\xspace}

 \def\Ppsi        {\ensuremath{\psi}\xspace}                 
                  
 \mathchardef\PDelta="7101
 \mathchardef\PXi="7104
 \mathchardef\PLambda="7103
 \mathchardef\PSigma="7106
 \mathchardef\POmega="710A
 \mathchardef\PUpsilon="7107
                  
 \def\PB      {\ensuremath{B}\xspace}                 
                  
 \def\PD      {\ensuremath{D}\xspace}

 \def\PJ      {\ensuremath{J}\xspace}                 
 \def\PK      {\ensuremath{K}\xspace}

 \def\Pc      {\ensuremath{c}\xspace}

 \def\Pi      {\ensuremath{i}\xspace}

 \def\Pp      {\ensuremath{p}\xspace}

}

\makeatletter
\ifcase \@ptsize \relax  \newcommand{\miniscule}{\@setfontsize\miniscule{4}{5}}\or  \newcommand{\miniscule}{\@setfontsize\miniscule{5}{6}}\or  \newcommand{\miniscule}{\@setfontsize\miniscule{5}{6}}\fi
\makeatother

\DeclareRobustCommand{\optbar}[1]{\shortstack{{\miniscule (\rule[.5ex]{1.25em}{.18mm})}
  \\ [-.7ex] $#1$}}

\def\cquark    {{\ensuremath{\Pc}}\xspace}

\def\pion   {{\ensuremath{\Ppi}}\xspace}

\def\pip    {{\ensuremath{\pion^+}}\xspace}

\def\pipm   {{\ensuremath{\pion^\pm}}\xspace}

\def\kaon    {{\ensuremath{\PK}}\xspace}
  \def\Kbar    {{\kern 0.2em\overline{\kern -0.2em \PK}{}}\xspace}

\def\KorKbar    {\kern 0.18em\optbar{\kern -0.18em K}{}\xspace}

\def\Km      {{\ensuremath{\kaon^-}}\xspace}

  \def\Dbar    {{\kern 0.2em\overline{\kern -0.2em \PD}{}}\xspace}
\def\D       {{\ensuremath{\PD}}\xspace}

\def\DorDbar    {\kern 0.18em\optbar{\kern -0.18em D}{}\xspace}
\def\Dz      {{\ensuremath{\D^0}}\xspace}

\def\Dstarpm {{\ensuremath{\D^{*\pm}}}\xspace}

\def\Bbar    {{\ensuremath{\kern 0.18em\overline{\kern -0.18em \PB}{}}}\xspace}

\def\BorBbar    {\kern 0.18em\optbar{\kern -0.18em B}{}\xspace}

\def\jpsi     {{\ensuremath{{\PJ\mskip -3mu/\mskip -2mu\Ppsi\mskip 2mu}}}\xspace}

  \def\Y#1S{\ensuremath{\PUpsilon{(#1S)}}\xspace}

\def\proton      {{\ensuremath{\Pp}}\xspace}

\def\Lz          {{\ensuremath{\PLambda}}\xspace}
\def\Lbar        {{\ensuremath{\kern 0.1em\overline{\kern -0.1em\PLambda}}}\xspace}
\def\LorLbar    {\kern 0.18em\optbar{\kern -0.18em \PLambda}{}\xspace}

\def\Lc      {{\ensuremath{\Lz^+_\cquark}}\xspace}

\newcommand{\decay}[2]{\ensuremath{#1\!\to #2}\xspace}         
\def\to                 {\ensuremath{\rightarrow}\xspace}

\def\AT#1     {\ensuremath{A_{\mathrm{T}}^{#1}}\xspace}

\def\C#1      {\ensuremath{\mathcal{C}_{#1}}\xspace}                       \def\Cp#1     {\ensuremath{\mathcal{C}_{#1}^{'}}\xspace}                    \def\Ceff#1   {\ensuremath{\mathcal{C}_{#1}^{\mathrm{(eff)}}}\xspace}        \def\Cpeff#1  {\ensuremath{\mathcal{C}_{#1}^{'\mathrm{(eff)}}}\xspace}       \def\Ope#1    {\ensuremath{\mathcal{O}_{#1}}\xspace}                       \def\Opep#1   {\ensuremath{\mathcal{O}_{#1}^{'}}\xspace}

\DeclareSIUnit\clight{\text{\ensuremath{c}}}
\DeclareSIUnit\micron{\micro\metre}
\DeclareSIUnit\mrad{\milli\radian}
\DeclareSIUnit\gauss{G}

\DeclareSIUnit\meV{\milli\eV}
\DeclareSIUnit\keV{\kilo\eV}
\DeclareSIUnit\MeV{\mega\eV}
\DeclareSIUnit\GeV{\giga\eV}
\DeclareSIUnit\TeV{\tera\eV}

\DeclareSIUnit[per-mode=symbol]\MeVc{\MeV\!\per\clight}
\DeclareSIUnit[per-mode=symbol]\GeVc{\GeV\!\per\clight}

\DeclareSIUnit[per-mode=symbol]\MeVcc{\MeV\!\per\clight\squared}
\DeclareSIUnit[per-mode=symbol]\GeVcc{\GeV\!\per\clight\squared}
\DeclareSIUnit[per-mode=symbol]\GeVGeVcccc{\GeV\squared\!\per\clight^{4}}

\DeclareSIUnit\kB{\kilo\byte}
\DeclareSIUnit\kHz{\kilo\hertz}
\DeclareSIUnit\MHz{\mega\hertz}
\DeclareSIUnit[per-mode=symbol]\MBs{\mega\byte\per\second}
\DeclareSIUnit[per-mode=symbol]\GBs{\giga\byte\per\second}

\DeclareSIUnit\mb{\micro\barn}
\DeclareSIUnit\nb{\nano\barn}
\DeclareSIUnit\pb{\pico\barn}
\DeclareSIUnit\fb{\femto\barn}
\DeclareSIUnit\ab{\atto\barn}
\DeclareSIUnit\zb{\zepto\barn}
\DeclareSIUnit\yb{\yocto\barn}

\DeclareSIUnit\invnb{\per\nano\barn}
\DeclareSIUnit\invpb{\per\pico\barn}
\DeclareSIUnit\invfb{\per\femto\barn}
\DeclareSIUnit\invab{\per\atto\barn}

\DeclareSIUnit\Xrad{\text{\ensuremath{X_{0}}}}
\DeclareSIUnit\NIL{\text{\ensuremath{\lambda_{\text{int}}}}}
\DeclareSIUnit\mip{MIP}

\def\gsim{{~\raise.15em\hbox{$>$}\kern-.85em
          \lower.35em\hbox{$\sim$}~}\xspace}
\def\lsim{{~\raise.15em\hbox{$<$}\kern-.85em
          \lower.35em\hbox{$\sim$}~}\xspace}

\def\tesla      {\mbox{\textsc{Tesla}}\xspace}

\def\gauss      {\mbox{\textsc{Gauss}}\xspace}

\def\cpp        {\mbox{\textsc{C\raisebox{0.1em}{{\footnotesize{++}}}}}\xspace}

\def\tell1  {TELL1\xspace}
\def\ukl1   {UKL1\xspace}

\usepackage[nameinlink,capitalise]{cleveref}

\usepackage{cite} \usepackage{mciteplus}
 \usepackage{longtable} 

\hypersetup{pdfauthor={\paperauthors},
            pdftitle={\paperasciititle},
            pdfkeywords={\paperkeywords},
            pdfcopyright={Copyright (C) \papercopyright},
            pdflicenseurl={\paperlicenceurl}}

\begin{document}

\renewcommand{\thefootnote}{\fnsymbol{footnote}}
\setcounter{footnote}{1}

\begin{titlepage}
\pagenumbering{roman}

\vspace*{-1.5cm}
\centerline{\large EUROPEAN ORGANIZATION FOR NUCLEAR RESEARCH (CERN)}
\vspace*{1.5cm}
\noindent
\begin{tabular*}{\linewidth}{lc@{\extracolsep{\fill}}r@{\extracolsep{0pt}}}
\ifthenelse{\boolean{pdflatex}}{\vspace*{-1.5cm}\mbox{\!\!\!\includegraphics[width=.14\textwidth]{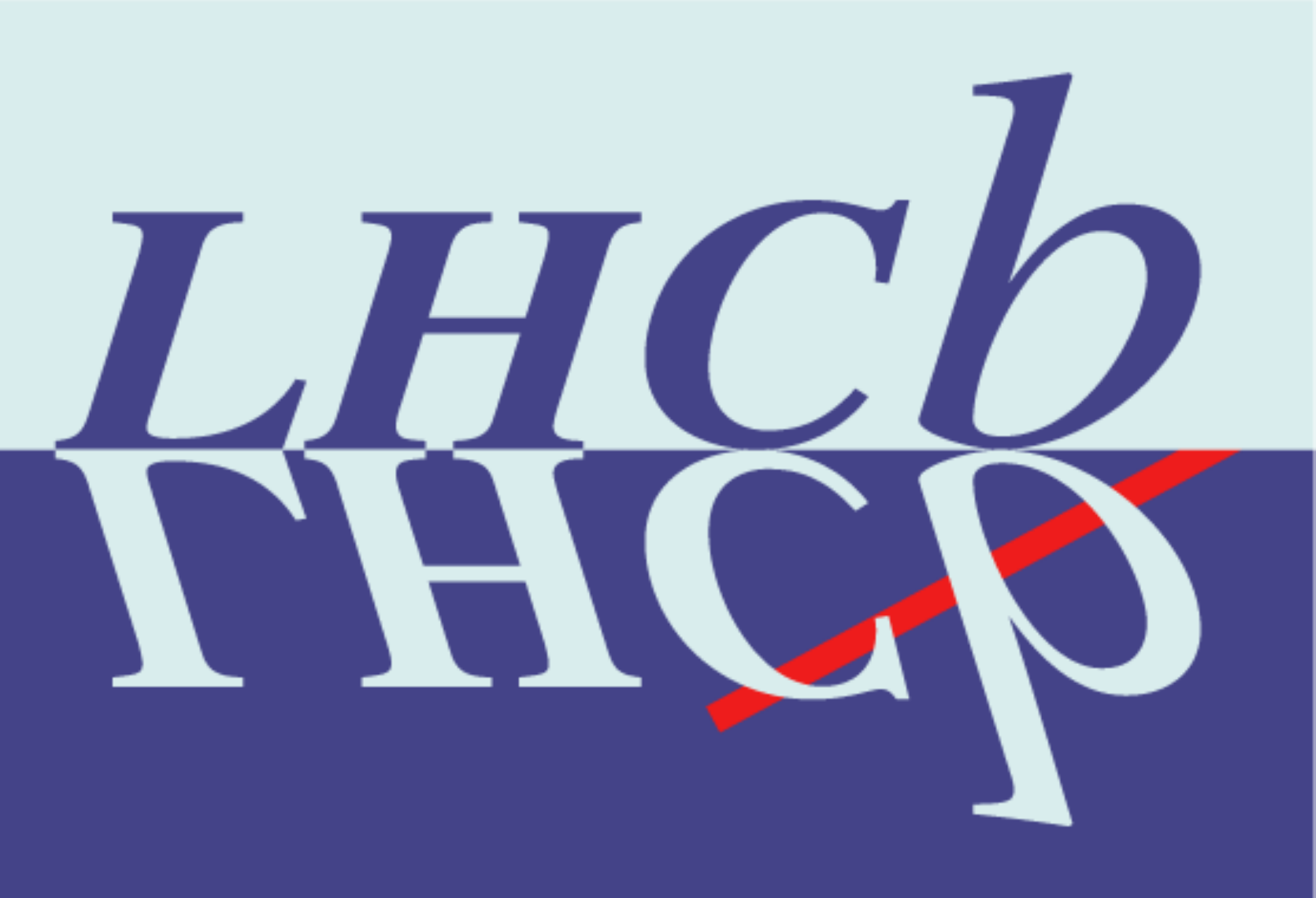}} & &}{\vspace*{-1.2cm}\mbox{\!\!\!\includegraphics[width=.12\textwidth]{lhcb-logo.eps}} & &}\\
 & & \\
 & & CERN-LHCb-DP-2019-002 \\   & & \today \\  & & \\
\end{tabular*}

\vspace*{2.0cm}

{\normalfont\bfseries\boldmath\huge
\begin{center}
  \papertitle 
\end{center}
}

\vspace*{0.75cm}

\author[1]{R.~Aaij}
\author[1]{S.~Benson}
\author[2]{M.~De~Cian}
\author[3]{A.~Dziurda}
\author[4]{C.~Fitzpatrick}
\author[1]{E.~Govorkova}
\author[5]{O.~Lupton}
\author[4]{R.~Matev}
\author[6]{S.~Neubert}
\author[4]{A.~Pearce}
\author[7]{H.~Schreiner}
\author[4]{S.~Stahl}
\author[5]{M.~Vesterinen}

\affil[1]{Nikhef National Institute for Subatomic Physics, Amsterdam, Netherlands}
\affil[2]{Institute of Physics, Ecole Polytechnique F\'ed\'erale de Lausanne (EPFL), Lausanne, Switzerland}
\affil[3]{Henryk Niewodniczanski Institute of Nuclear Physics Polish Academy of Sciences, Krak\'ow, Poland}
\affil[4]{European Organization for Nuclear Research (CERN), Geneva, Switzerland}
\affil[5]{Department of Physics, University of Warwick, Coventry, United Kingdom}
\affil[6]{Physikalisches Institut, Ruprecht-Karls-Universit\"at Heidelberg, Heidelberg, Germany}
\affil[7]{University of Cincinnati, Cincinnati, OH, United States}

\begin{center}
  \makeatletter\@author\makeatother
\end{center}

\vspace*{0.75cm}

\begin{abstract}
  \noindent
  An evolved real-time data processing strategy is proposed for high-energy 
  physics experiments, and its implementation at the LHCb experiment is 
  presented.
  The reduced event model allows not only the signal candidate firing the 
  trigger to be persisted, as previously available, but also an arbitrary set 
  of other reconstructed or raw objects from the event.
  This allows for higher trigger rates for a given output data bandwidth, when 
  compared to the traditional model of saving the full raw detector data for 
  each trigger, whilst accommodating inclusive triggers and preserving data 
  mining capabilities.
  The gains in physics reach and savings in computing resources already made 
  possible by the model are discussed, along with the prospects of employing it 
  more widely for Run 3 of the Large Hadron Collider.
\end{abstract}

\vspace*{1.0cm}

\begin{center}
  Published in JINST 14 P04006
\end{center}

\vspace{\fill}

{\footnotesize 
\centerline{\copyright~\papercopyright, licence \href{\paperlicenceurl}{\paperlicence}.}}
\vspace*{2mm}

\end{titlepage}

\newpage
\setcounter{page}{2}
\mbox{~}

\cleardoublepage

\renewcommand{\thefootnote}{\arabic{footnote}}
\setcounter{footnote}{0}

\pagestyle{plain} \setcounter{page}{1}
\pagenumbering{arabic}

\section{Introduction}
\label{sec:intro}

Experimental tests of the Standard Model must become ever more precise if small
effects due to new physics are to be observed.
To meet this challenge, at the \ac{LHC} both the centre-of-mass energy of the
colliding proton-proton beams and the instantaneous luminosity delivered to the
experiments are periodically increased.
The corresponding increase in signal production rate must be balanced against
the availability of computational resources required to store the data for
offline analysis.
The disk space required is given by the product of the running time of the
experiment and the trigger output bandwidth defined as
\begin{equation*}
  \text{Bandwidth}\,[\si{\MBs}] \propto \text{Trigger output rate}\,[\si{\kilo\hertz}] \times \text{Average event size}\,[\si{\kilo\byte}].
\end{equation*}
When the output rate of any given trigger is dominated by events containing
signal processes, tightening the selection further to reduce the output rate is
undesirable.
The size on disk of the full raw detector information cannot be decreased
beyond the standard zero-suppression and compression techniques. Therefore,  reduced event
formats must be employed instead, wherein a subset of the high-level
reconstructed event information computed in the final software trigger stage is
recorded and sent to permanent storage.
As the rate of signal processes increases, so must the reliance on reduced
event formats.

The \cms, \lhcb, and \atlas experiments utilised their own versions of reduced
formats during
\runtwo (2015--2018)~\cite{CMS:2012ooa,LHCb-DP-2016-001,ATL-DAQ-PUB-2017-003}.
Typically, the reduced format contains information pertaining only to the
reconstructed physics objects which passed the relevant set of trigger selections, as
well as some event summary information.
For a jet trigger, for example, the objects may be the momentum vector and
particle multiplicity value of the highest energy jet in an event. For
a heavy flavour decay trigger it may be a set of four momenta and decay vertex
positions.
Such an approach provides a maximal reduction in persisted event size whilst still
allowing for the analysis performed in the trigger to be continued offline.
This allows for higher trigger rates for a given output bandwidth, extending
the physics reach of an experiment within limited computational resources.
However, it also restricts the utility of an event for broader analysis and
data mining.
Such a reduction is then unsuitable for inclusive triggers, which constitute a
large fraction of trigger output rates, as well as for analyses in which other
information may be required later, such as performing flavour tagging or
isolation studies, the exact input to which is not well-defined at the time of
the trigger decision.

The \lhcb experiment has pioneered the widespread usage of a reduced event format
since the beginning of \runtwo.
This has been driven by the desire to continue the rich charm physics programme
started in \runone~(2010--2012) in spite of the large charm production rate in the detector
acceptance, being 25 times larger than that for
beauty~\cite{LHCb-PAPER-2015-037,LHCb-PAPER-2015-041}.
During \runtwo, almost all events selected by charm triggers at \lhcb were persisted in a reduced
format, enabling a broader physics programme which would otherwise not be possible within the available
resources.
In \runthree (2021--2023) the instantaneous luminosity delivered to the \lhcb experiment will increase by a factor of five.
The rate of events processed by the trigger that contain charm or beauty hadrons will scale not
only by this factor, but also by a
factor of two due to the implementation of a higher-efficiency full software
trigger~\cite{LHCb-TDR-012,LHCb-TDR-016}.
Coupled with the larger raw data size produced by the upgraded detector, 
it is necessary for some fraction of the beauty programme to migrate to reduced event formats in 
order to fit within available computational resources.
Given the strong reliance the beauty programme has on inclusive
triggers~\cite{LHCb-DP-2012-004,LHCb-DP-2019-001}, the style of reduced event formats previously
described are not sufficient, and so this model must be extended if
the charm and beauty programmes are to be sustained into the future.

A significant evolution of the reduced event model is proposed here, in which
additional reconstructed objects in the event are selected and persisted after
the trigger decision has been made.
When implemented with sufficient flexibility, this allows for fine-grained
tuning between trigger output bandwidth and event utility offline.
Since mid-2017, such a scheme has been adopted in the \lhcb trigger, resulting
in substantial bandwidth savings without a loss of physics reach.
The reduction in bandwidth has allowed for the introduction of new trigger
selections, increasing the physics scope of the experiment.
It is foreseen that the reduced event format will be adopted for a majority of
the \lhcb physics programme in \runthree~\cite{LHCb-TDR-018}.
The latest evolution, presented here, now caters to all use cases.

The rest of this paper is organised as follows.
The \lhcb detector and trigger system is described in \cref{sec:trigger}, and
the need for a fully flexible reduced event format in \runtwo is
motivated quantitatively.
A technical overview of the new format is given in \cref{sec:turbo}.
The benefits already achieved in \runtwo and the prospects for \runthree are presented in \cref{sec:gains,sec:prospects}.
A summary is given in \cref{sec:summary}.
 \section{The LHCb detector and trigger strategy}
\label{sec:trigger}

The \lhcb detector is a forward-arm spectrometer designed to measure the
production and decay properties of charm and beauty hadrons with high
precision~\cite{Alves:2008zz,LHCb-DP-2014-002}.
Such objects are predominantly produced at small angles with respect to the
proton beam axis~\cite{LHCb-PAPER-2010-002}.
A tracking system is used to reconstruct the trajectories of charged particles.
It consists of a silicon vertex detector surrounding the interaction
region~(\velo), a silicon strip detector located upstream of a dipole magnet
(\ttracker), and three tracking stations downstream of the magnet. The latter
each consist of a silicon strip detector in the high-intensity region close to the
beamline (\intr) and a straw-tube tracker in the regions further from the
beamline (\ot).
Neutral particles are identified with a calorimeter system made of a
scintillating pad detector (\spd), an electromagnetic calorimeter (\ecal) preceded by a
pre-shower detector (\presh), and a hadronic calorimeter (\hcal).
Charged particle identification is provided by combining information from the
ring-imaging Cherenkov detectors (\richone and \richtwo), wire chambers
used to detect muons, and the calorimeter system.
Shower counters located in high-rapidity regions either side of the main
detector can be used to veto events with particles produced at a very low angle
to the beam, mainly used for studies of central exclusive
production~\cite{LHCb-DP-2016-003}.
The \ac{LHC} provides proton-proton collisions to the experiment at a rate of
\SI{30}{\mega\hertz} during most run periods.
The detector also operates during special physics runs, such as heavy
ion collisions with lead and xenon nuclei, and in a fixed target mode where a
noble gas is injected in the interaction
region~\cite{Barschel:2014iua,LHCb-PAPER-2014-047}.

\begin{figure}
  \begin{center}
    \includegraphics{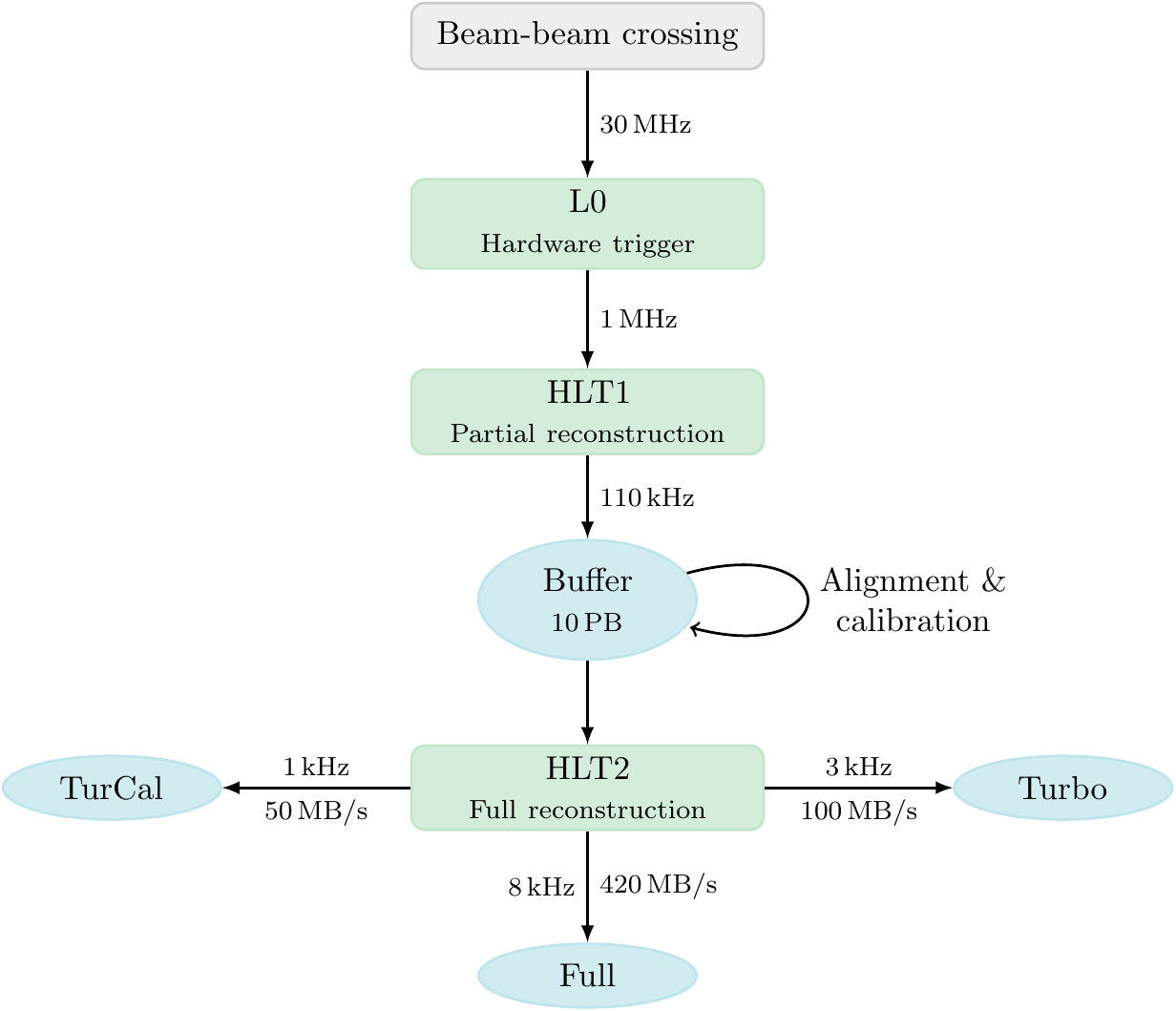}
  \end{center}
  \caption{    An overview of the LHCb trigger scheme in \runtwo~\cite{LHCb-DP-2019-001}.
    The green boxes represent trigger stages, the blue ellipses represent
    storage units, and arrows between elements represent data flow, labelled
    with approximate flow rates.
    Events that reach a terminal storage unit are kept for offline analysis.
  }
  \label{fig:trigger_scheme}
\end{figure}

During proton-proton operation of the \ac{LHC}, an interesting physics signal process occurs at a
rate of around \SI{10}{\hertz}.
To filter out the large majority of collisions that do not contain interesting
information, and to fit the experimental output rate into the available
computing resources, a three-stage trigger system is employed~\cite{LHCb-DP-2019-001},  illustrated in
\cref{fig:trigger_scheme}.
In total, it reduces the event rate by three orders of magnitude, with the
output being sent to permanent storage.
The level-0 hardware trigger~(\lzero) uses information from the calorimeter and
muon systems to compute a decision using field-programmable gate arrays within
a fixed latency of \SI{4}{\micro\second}.
Events are selected by the \lzero at a rate of about \SI{1}{\MHz}, and are sent
to an \ac{EFF} where they are processed by the two-stage High Level
Trigger~(\hlt) on commodity processors.
The first stage, \hltone, uses tracking and calorimetry information to perform
a partial reconstruction of charged particles, and writes the raw information
for each passing event to a \SI{10}{\peta\byte} disk buffer at a rate of around
\SI{110}{\kHz}.
Asynchronously to \hltone, the data in the buffer are used to compute alignment
and calibration constants which, if significantly different from previous runs,
are saved to a conditions database~\cite{Dujany:2017839}.
The second stage of the software trigger, \hlttwo, performs a full event
reconstruction using the latest alignment and calibration constants and all
available detector information.
Selections in \hlttwo span a spectrum from \emph{inclusive} selections, which
require the presence of a heavy flavour decay signature such as a displaced
multi-body vertex or a high transverse momentum lepton, and \emph{exclusive}
selections, which fully reconstruct signal decays.
Events are written to offline storage from \hlttwo at a rate of around
\SI{12.5}{\kHz}.

Within the \hlt, an event comprises a set of so-called sub-detector raw banks,
each containing the zero-suppressed readout of a given sub-detector.
Each trigger stage adds a set of trigger raw banks to the event which summarise
how the event passed that stage.
Selected events are persisted to a set of \emph{streams} in permanent storage.
During this \emph{streaming}, different raw banks may be kept or removed
depending on the stream the event is sent to.
Each trigger selection in \hlttwo is associated to a particular stream, and an 
event is sent to a stream if it passes at least one associated selection.
An event may be sent to multiple streams, and different sets of raw banks for
that event can be saved in each stream.
The \lhcb physics trigger rate is distributed over three streams:
\begin{enumerate}
 \item Around half of all triggered events enter the \emph{full} stream,
 which contains the full set of sub-detector and trigger raw banks, with the
 trigger reconstruction being discarded.
 Events in the full stream are reconstructed by a separate application offline,
 followed by a set of physics selections that reconstruct thousands of decay
 chains for direct analysis.
 This scheme is used in many other experiments, where it is common for the trigger
 reconstruction to be of poorer quality to that of the offline reconstruction
 in order to fit within the timing constraints of their online systems.
 \item Around a third of triggered events are sent to the \emph{Turbo} stream, where a reduced
 event format is persisted in a dedicated raw bank, described in detail in \cref{sec:turbo}.
 \item The remaining trigger rate is captured by the \emph{TurCal} calibration
   stream, where both the reduced and full formats are kept.
\end{enumerate}

The motivation for creating the reduced format used in the Turbo and TurCal streams now follows.

\subsection{Real-time analysis}

Here, `real time' is defined as the interval between a collision occurring and
the point at which the corresponding event must be either discarded forever or
sent offline for permanent storage.
In most high-energy physics experiments, this interval cannot exceed the time
between collisions as their trigger systems are synchronous.
A multi-stage trigger increases the available event processing time from one stage to the next;
however each stage requires more computing power than the previous one.
The addition of a disk buffer between two stages increases the effective
computation time allowed in the latter stage as that stage can then also
operate during periods when data are not being taken.
The buffer also permits the execution of tasks that provide additional input to
the latter trigger stage, such as the running of alignment and calibration
algorithms.

During \runone, \lhcb installed a disk buffer such
that \SI{20}{\percent} of the \lzero output was deferred to disk.
This increased the effective processing power of the software trigger by 
allowing it to run both during beam operation and when the \ac{LHC} was in 
downtime.
Between \runone and \runtwo, the buffer size was increased to \SI{10}{\peta\byte},
with events being buffered between the two software trigger stages.
As the buffer is so large, `real time' can be up to two weeks during nominal 
data-taking~\cite{LHCb-DP-2019-001}.
The possibility to buffer the data for such a long period has allowed for the 
execution of the aforementioned real-time detector alignment and calibration on 
the data in the buffer. The increase in computing power has permitted the 
implementation of the full offline reconstruction in \hlttwo.
With these substantial additions, the \hlttwo reconstruction is of equal
quality to what is achieved in the offline processing, such that full
physics selections are performed in real time in the trigger without a loss of
precision.
This permits trigger selections to be much closer or identical to those applied
in the offline analysis, as there are no resolution effects between online and
offline to be accounted for.
Such a scheme reduces experimental systematic uncertainties and saves money by making
tighter trigger selections acceptable, therefore reducing the rate and output bandwidth.

With an offline-quality reconstruction in the final trigger stage~(\hlttwo),
it is no longer necessary to run another reconstruction offline.
Instead, the objects created by trigger selections are written out to the permanent
storage directly.
Physics measurements are performed on these objects.
This technique almost eliminates processing requirements offline and reduces output
bandwidth, if the relevant subset of the reconstruction is smaller than the raw
event.
Offline analysis of trigger-level information has been used at the \cms
experiment~\cite{CMS:2012ooa}, called `scouting', and at the \atlas
experiment~\cite{ATL-DAQ-PUB-2017-003}, called `trigger-object level analysis'.
Although the method has extended their physics reach, in both cases only object
summaries are available for analysis, rather than offline-equivalent
constructs, and the quality of the reconstruction is worse than that achieved
offline.
Since 2015, the \lhcb experiment has employed a persistence model called
`Turbo'. It allows the offline-equivalent information computed in \hlttwo to be saved, sacrificing
neither physics performance nor analyst convenience, as existing tools are
used to process the data.

Until recently, the reduced event formats employed by the \ac{LHC} experiments
have described only the objects that enter into some trigger selection, such as
a jet or a charm hadron and its decay products.
These formats are best used to reduce the output bandwidth of
events captured by exclusive trigger selections.
In contrast, an inclusive trigger selection, which does
not necessarily consider all the information on the physics process of
interest in its decision, cannot be accommodated.
In order to be able to cater for such use cases, an advanced reduced 
persistency model has recently been developed as an evolution of the Turbo 
model, and is described in the following \namecref{sec:turbo}.
 \section{The Turbo data processing model}
\label{sec:turbo}

The Turbo data processing model has evolved considerably over the course of \runtwo. The
initial prototype, established in 2015 and described in detail in
Ref.~\cite{LHCb-DP-2016-001}, saved the set of objects associated to individual reconstructed decay cascades extracted from events fully reconstructed in the trigger.
This exploits the event topology characteristic of hadron colliders wherein only a small subset of objects produced in the collision are relevant for analysis offline.
In a hard proton-proton scatter at \lhcb, on average 40 tracks are associated 
to the resulting primary vertex, whereas only 2--6 tracks are required to 
reconstruct a typical heavy flavour decay.
Significant savings in persisted event size are then possible by discarding reconstructed objects not needed in the offline analysis.
In this
section, three new developments to the Turbo processing model are described,
which together allow for all \lhcb analyses to be accommodated.

\begin{figure}
  \begin{center}
    \includegraphics{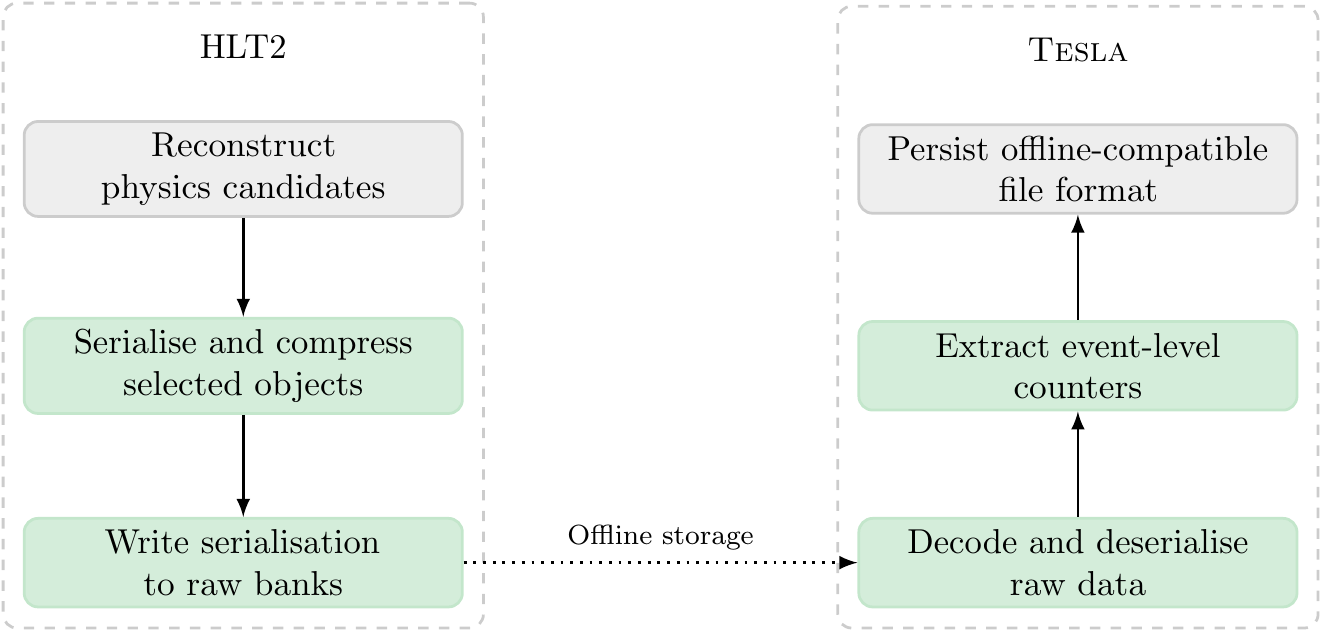}
    \caption{      Data flow in saving and restoring trigger objects from online (left) to
      offline (right)~\cite{LHCb-DP-2016-001}.
    }
    \label{fig:reports}
  \end{center}
\end{figure}

In order to perform physics analyses with the output of the trigger reconstruction,
decay candidates must appear in the same format used by existing analysis tools.
Containers of physics object classes are serialised per event into raw banks, as illustrated in \cref{fig:reports}, in order to conform to the trigger output format.
This format is optimised for simple event-by-event concatenation, rather than the heavily compressed format used offline.
The initial prototype of the Turbo model~\cite{LHCb-DP-2016-001}
serialised only the candidates that enter the trigger decision.
To allow for additional objects to be persisted, the serialisation framework
used in the offline infrastructure was adapted to work within the online system.
This increases the compatibility between analyses using the online and offline reconstructions,
and requires that only one serialisation framework be maintained.
Furthermore, trigger selection configurations which use the Turbo model were extended to allow for a set of `additional selections' 
to be run \emph{after} the trigger decision has been made. These allow for any reconstructed object to be captured.
At the end of each event processing, the set of all \cpp physics objects
selected by all trigger lines using the Turbo model is copied to a common location in memory.
The instances in the copied location are
then compressed and serialised into raw banks~\cite{Albrecht:2310579}, suitable for transfer within and out of the online system.

A dedicated application, called \tesla~\cite{LHCb-DP-2016-001}, runs offline
to transform the \hlttwo output
into a format that is ready for analysis. This involves a file conversion from
serialised raw data to a more compressed format used by the offline storage, as well as
the computation and storage of information necessary for luminosity determination.
\tesla also ensures that additional information calculated in the online reconstruction is accessible to standard analysis tools, for example
event-level sub-detector occupancies and information calculated using the whole reconstructed event.
When processing simulated events, the application also matches reconstructed objects to
true simulated objects and stores the information as relations tables.
In comparison with the traditional model of an additional offline
reconstruction, the processing cost of running \tesla is negligible.
In principle, some or even all parts of the work done by \tesla can
be moved into the trigger itself in the \lhcb upgrade, such as if
compressed object containers were written directly out of \hlttwo instead of encoding them into raw banks.

With the serialisation and file preparation frameworks in place, different levels of granularity on what physics objects to select are now available.
In the following, the resulting flexibility is explained, such that all
measurements can take advantage of the real-time model.

\subsection{Standard Turbo model}
\label{sec:turbo:standard}

A majority of trigger lines based on the Turbo model define exclusive selections where the full decay is
completely specified, and no additional objects from the event are required for subsequent analysis.
In this model, the objects saved are:
\begin{itemize}
  \item The reconstructed decay chain that fired the line, which comprises:
    \begin{itemize}
      \item The set of all tracks and neutral objects, calorimeter and PID
        information relating to those objects, and decay vertices that form the
        candidate.
      \item The tracking detector clusters associated to the candidate tracks, such that the
        tracks can be re-fitted offline.
    \end{itemize}
  \item All of the reconstructed \acp{PV} in the event, which are necessary to
    perform \ac{PV} mis-association studies offline.
\end{itemize}
Other reconstructed objects in the event as well as the raw
data are not kept for offline processing. That allows for a significant
reduction of the event size and hence also trigger output bandwidth.
Offline disk space and CPU processing time are also saved as the offline reconstruction step is omitted.
This model has been operational since the beginning of data-taking in 2015, and
enabled the first \lhcb \runtwo measurements to be presented 18~days after the
data were collected~\cite{LHCb-PAPER-2015-037,LHCb-PAPER-2015-041}.

\subsection{Complete reconstruction persistence}
\label{sec:turbo:complete}

In order to use the Turbo model for inclusive triggers, the ability to store all the
reconstructed objects in the event was introduced in the beginning of 2016.
It is made available on a per-selection basis by a user flag.
When enabled, the full event reconstruction as performed in \hlttwo is persisted
in addition to the information described in \cref{sec:turbo:standard}.
In comparison to saving the raw event, this approach reduces disk space usage and 
requires no further processing offline. However, the information needed to 
re-run the reconstruction offline is discarded.
This technique permits high-rate inclusive triggers that would otherwise not be 
feasible~\cite{LHCb-PAPER-2016-064}.

Persisting the whole reconstructed event is expensive in terms of event size in
comparison with only saving the trigger candidate, as shown in
\cref{tab:event_size}.
\begin{table}
  \caption{    Average event sizes for trigger lines requesting varying levels of
    information persistence, measured on data collected in 2018.
  }
  \label{tab:event_size}
  \begin{center}
    \begin{tabular}{rS[table-format=1]}
  \toprule
  Persistence method & {Average event size (\si{\kB})}\\
  \midrule
  Turbo & 7\\
  Selective persistence & 16\\
  Complete persistence & 48\\
  Raw event & 69\\
  \bottomrule
\end{tabular}
   \end{center}
\end{table}
In most cases, only a small fraction of the full reconstructed event is
required in an offline analysis.
Therefore, a middle-ground between persisting only the candidate or the whole
reconstructed event is introduced.

\subsection{Selective reconstruction persistence}
\label{sec:turbo:sel_reco}

Selective persistence allows for
explicit specification of which information is stored on top of the trigger
candidate itself. This permits a significant event size reduction
without the usual sacrifice of allowing for exploratory analysis offline.
At the beginning of 2017, trigger lines were augmented with additional selections 
that are executed after the
trigger decision has been computed.
A typical additional selection captures objects in the event that are somehow
related to the trigger candidate. However, any selections are possible.
\begin{figure}
  \begin{center}
    \includegraphics[width=\textwidth]{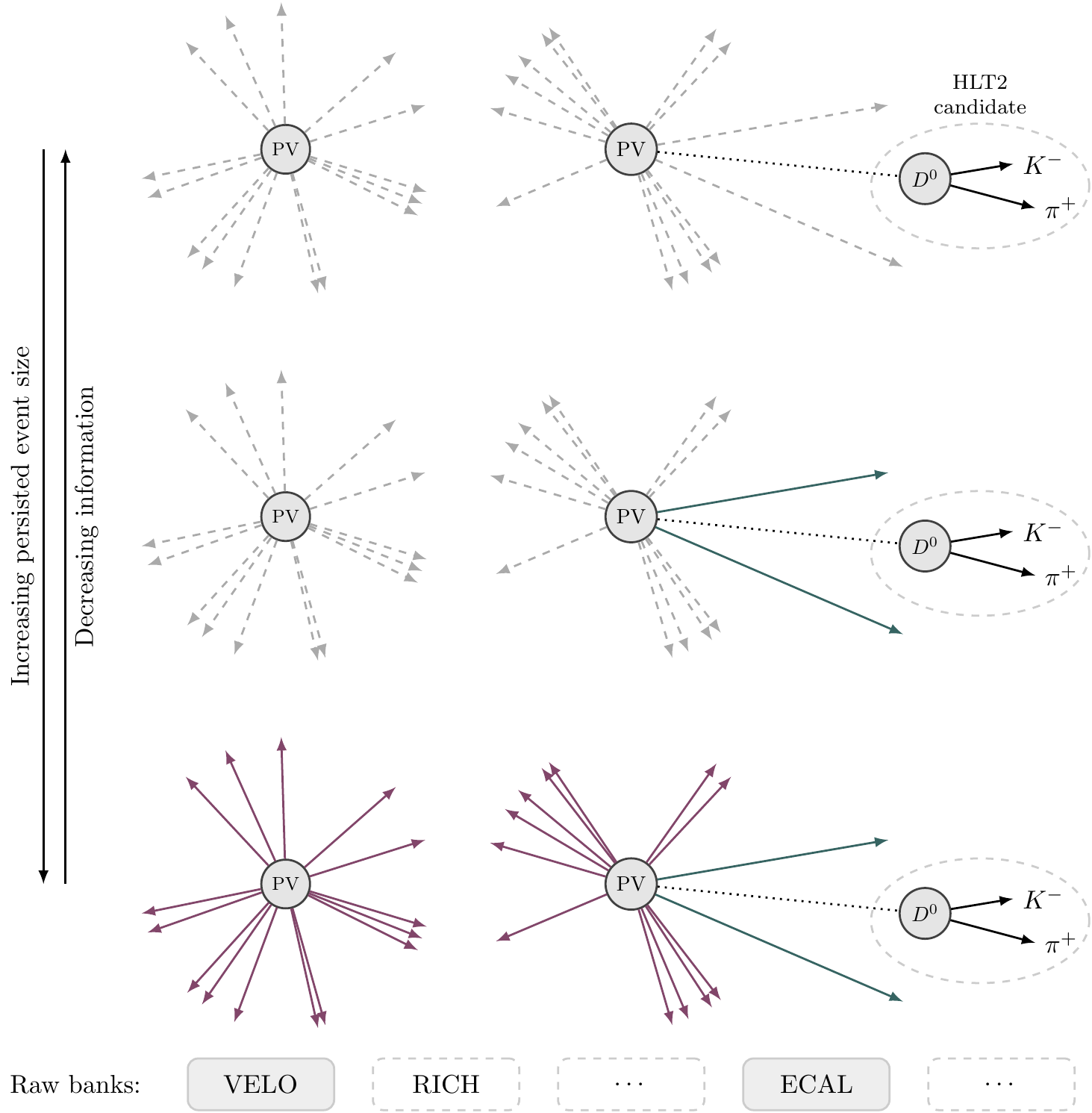}
    \caption{      A cartoon of the same reconstructed event with varying levels of object
      persistence: Turbo (top); selective persistence (middle); and complete
      reconstruction persistence (bottom).
      Solid objects are those persisted in each case.
      A trigger selection may also ask for one or more sub-detector raw 
      banks to also be stored, shown as solid rectangles.
    }
    \label{fig:event_persistency}
  \end{center}
\end{figure}

As an example, consider a trigger using the Turbo model that reconstructs and selects the \decay{\Dz}{\Km\pip} charm decay, as shown in \cref{fig:event_persistency}, where the information specified in \cref{sec:turbo:standard} is saved.
If complete reconstruction persistence is enabled for this line, the
underlying reconstructed event will also be stored. With selective persistence, additional objects are instead specified explicitly, such as all charged pions that are associated to the
same \ac{PV} as the \Dz and that form a good-quality excited \Dstarpm candidate.
The selection framework allows for requirements to be
made on both the pions themselves and on the $\Dz\pipm$ combination, but then only the pions that pass these cuts are persisted. The \Dstarpm candidates
are discarded, since it can be built again exactly in an offline processing if required.
Similar selections can be added for other extra particles, such as kaons,
photons, and hyperons to support a wide spectrum of charm spectroscopy
measurements with a single trigger selection.

In the \lhcb upgrade for \runthree, selective persistence is a key ingredient in
the migration of the physics programme to the real-time analysis model~\cite{LHCb-TDR-018}.
One use-case is flavour tagging, a determination of the initial flavour
of a beauty or anti-beauty meson.
The decision of one or more tagging algorithms together with the probability of
the assigned flavour being wrong is computed using a set of reconstructed
objects in the event, namely objects that are related to the same \ac{PV} as
the signal decay.
This set can be loosely defined upfront and added as additional selections to
trigger lines which reconstruct the signal beauty hadron decays of interest.
As the tagging algorithms undergo improvements during data-taking, the flavour
tagging can be re-run offline using the information captured by the additional
selections.
Initial studies show that the set of reconstructed objects required as input to
the tagging algorithms constitute only around \SI{10}{\percent} of the space
that would be required for persisting the full reconstruction.

\subsection{Selective raw persistence}
\label{sec:turbo:sel_raw}

There are some cases in which saving all possible information from an event
is required. One important example is for efficiency measurements on calibration samples. Today,
detector-level efficiencies are determined from control channels whose trigger lines save the full raw detector information.
The high bandwidth associated with doing this means that the choice and selection of control channels is severely restricted.
This impacts the calibration sample size and leads to larger uncertainties on the efficiencies.
The ability to study the efficiency of a given aspect of the reconstruction requires saving detector hits
that were not used in the reconstructed object. However, the efficiency of reconstructing
tracks with the \ot, for example, would require saving only the raw banks associated with that sub-detector, and
would not need the banks associated with the calorimeter or the \rich sub-detectors. If a more fine-grained, per-trigger specification of required raw banks was possible, the resulting bandwidth savings would allow
the calibration samples to more than double in size, resulting in more precise efficiency determinations and therefore 
more accurate physics measurements.

\begin{figure}
  \begin{center}
    \includegraphics{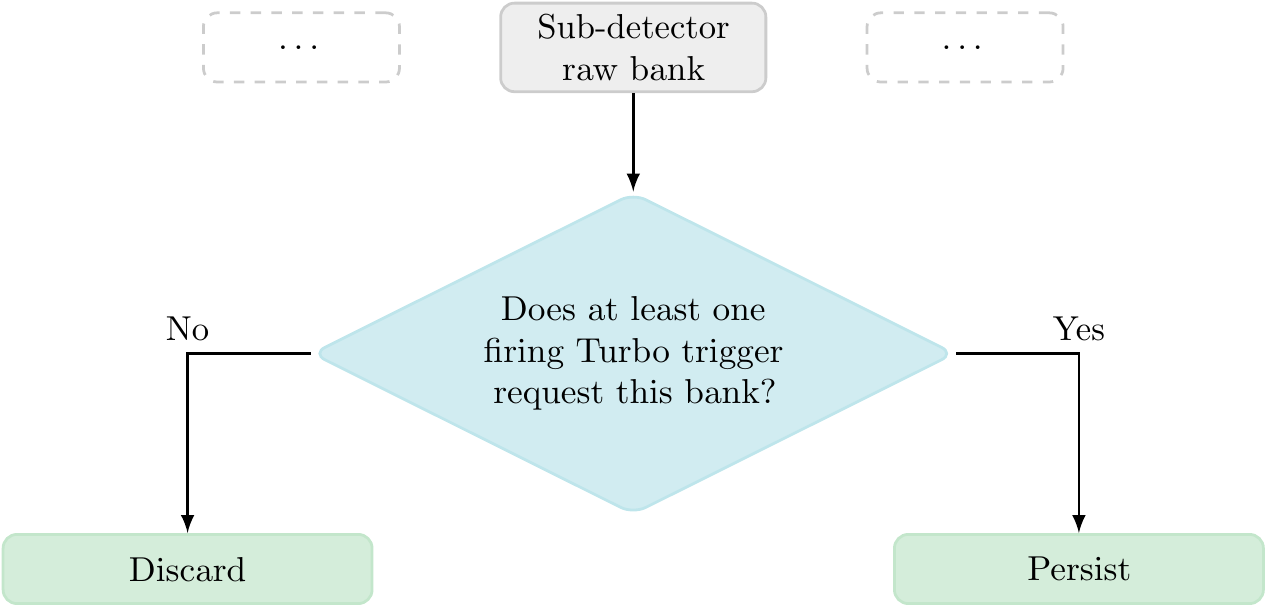}
    \caption{      Algorithm flow to decide whether a given raw bank is persisted for the current event. The resulting list of banks persisted
      for a given event is the superset of those required by the individual lines.
    }
    \label{fig:selraw}
  \end{center}
\end{figure}

In order to accomplish this, a new algorithm has been developed and deployed. On initialisation, it determines the list
of trigger selections and their requested raw banks from the trigger configuration,
as shown in \cref{fig:selraw}.
On a per-event basis, the decision of each trigger selection is examined and the superset of the required
raw banks is persisted to the corresponding output stream. Raw banks not requested by any firing trigger line are discarded.
This algorithm has been running in the \lhcb trigger since 2018.
With it, each trigger selection has complete control on additional information in the event is persisted, from any object created in the reconstruction, to the raw information created by any sub-detector.
Therefore, the Turbo model is now able to cater for any use case.
 \section{Achievements in \runtwo}
\label{sec:gains}

Since the beginning of \runtwo, a considerable fraction of triggered
events have used the Turbo model.
In 2018, the number of trigger lines using the Turbo model was around \SI{50}{\percent} of the
number using the traditional model, however the Turbo stream bandwidth was
\SI{25}{\percent} that of the full stream due to the reduced average event size.
Physics measurements using the Turbo model include charm and
\jpsi cross-sections~\cite{LHCb-PAPER-2015-037,LHCb-PAPER-2015-041}, the
discovery of new ground-state and excited charm
baryons~\cite{LHCb-PAPER-2017-002,LHCb-PAPER-2017-018}, searches for dark
photons~\cite{LHCb-PAPER-2017-038}, and the characterisation of charmonium production
within jets~\cite{LHCb-PAPER-2016-064}.
This \namecref{sec:gains} summarises how the new Turbo model has
increased physics reach whilst reducing the event size since its introduction
in 2016.

As an example of the gains that can and have been made, we consider a subset of
trigger lines using the Turbo model that exclusively reconstruct and select
Cabibbo-favoured decays of ground-state charm hadrons, such as
\decay{\Dz}{\Km\pip} and \decay{\Lc}{\proton\Km\pip} and their charge
conjugates, originating directly from beam-beam interactions.
These trigger selections are intended for calibration studies, as the properties of
these objects and decays are well known, as well as for charm spectroscopy, searching
for and characterising excited charm states that cascade down to the ground
states.
In 2016, these trigger lines were enhanced by the addition of the complete
reconstruction persistence, to allow for
excited states to be reconstructed offline, and
so the average event size increased from \SI{7}{\kB} in 2015 to \SI{48}{\kB}.\footnotemark
\footnotetext{  The average size in 2016 across all events using the Turbo model was
  \SI{42}{\kB}, illustrating the dominating rate of these selected processes
}
In 2017, the spectroscopy lines moved to the selective reconstruction
persistence, reducing the average event
size to \SI{16}{\kB}.
In turn, the bandwidth used by the Turbo stream decreased from \SI{139}{\MBs}
to \SI{79}{\MBs}.
The newly-available bandwidth was then utilised for new inclusive charm baryon
trigger lines that would have otherwise not been possible within the given
computational resources.
The selections used to reduce the set of persisted objects from the
spectroscopy lines were aligned with the ones used for
spectroscopy offline.
The additional selections are then \SI{100}{\percent} efficient with respect to the offline selections by definition.
 \section{Prospects for \runthree}
\label{sec:prospects}

The \lhcb detector will be upgraded for \runthree, where the
instantaneous luminosity will increase from that in \runtwo by a factor of
five.
With the corresponding increase in heavy flavour production rate, in addition
to expected trigger efficiency gains and an increase in the raw event size, the
trigger output bandwidth will increase by an order of magnitude from Run 2.
Due to constraints on offline storage resources, the bandwidth is limited to a
maximum of \SI{10}{\GBs}~\cite{LHCb-TDR-018}.
The breadth of the current experimental physics programme can only survive
within these resources if the fraction of trigger output rate sent to the Turbo
stream increases by over a factor of two.
The flexibility of the reduced event model described here has been designed to
reach that goal without an inherent loss of physics performance or reach.
This \namecref{sec:prospects} briefly discusses some relevant techniques.

In principle, reducing the amount of persisted information comes with risk, as
the set of information needed for a given measurement is not always known
upfront, and indeed the analysis itself has often not been conceived.
Information discarded in the trigger can be later required by
some unforeseen analysis offline.
However, the factor-five increase in instantaneous luminosity means a
corresponding increase in the average number of visible beam-beam interactions
per bunch crossing. Therefore, a relatively safe selection is to discard objects
which are identified as originating from primary vertices other than those
associated to the signal trigger object.
This information is not relevant to an analysis of the trigger object as it is
unrelated to the signal process, having been produced from independent
parton-parton scatters.\footnotemark
\footnotetext{Such a technique is similar in concept to jet grooming, in that
one wishes to keep only the objects associated to the hard scatter which
produced the heavy flavour quark-antiquark pair present in the event.}
Given a fully reconstructed signal candidate, its associated primary vertex can
be defined as the one with the smallest impact parameter with respect to the
signal momentum vector.
Primary vertices from which associated information should not be persisted can
then be identified using a minimum impact parameter cut, the exact value of which
can be tuned based on the expected resolution available to distinguish separate
primary vertices.

Inclusive trigger selections of heavy flavour decays present a particularly
challenging event size reduction problem as they have very high rates.
The potentially incomplete reconstruction of the signal momentum vector reduces the accuracy of non-signal primary vertex suppression.
Complimentary selections that reduce the persisted information in inclusively triggered events include:
\begin{itemize}
  \item The rejection of tracks that form a very poor quality vertex with the
    inclusive candidate, as such tracks would not be used in an offline
    analysis; and
  \item The rejection of objects identified by a multivariate algorithm trained
    to distinguish uninteresting objects from those associated to the portion
    of the signal process captured by the inclusive selection.
\end{itemize}
Preliminary studies have shown such techniques can fully capture all signal
objects in an inclusively triggered event with an efficiency of around
\SI{90}{\percent}, compared to saving the full reconstruction, whilst rejecting
over \SI{90}{\percent} of the unrelated objects.
This reduces the bandwidth of the prototype inclusive beauty trigger under study from
\SI{7.5}{\GBs} to \SI{1}{\GBs}.
While promising, further work is needed to quantify any possible biases which
may be associated to the signal decays which are not completely captured by the
selective persistence.
 \section{Summary and conclusions}
\label{sec:summary}

Traditional reduced event formats allow for a broader physics programme within
available computational resources, but this is usually countered by a poorer
quality reconstruction in the trigger and reduced data mining capabilities.
Since 2015, a real-time alignment and calibration procedure 
between trigger stages has allowed the \lhcb experiment
to deploy and exploit its Turbo model.
With this, offline-quality signal candidates are
persisted directly from the trigger for later analysis.
This has been crucial for the charm programme, which otherwise would have had
to significantly compromise its reach and diversity.

Since 2017, the implementation of the Turbo model has been overhauled and extended to allow an
arbitrary subset of the trigger reconstruction and raw sub-detector information
to be persisted along with the trigger candidate.
As such, the model is now capable of supporting the entirety of the
experiment's broad research programme, and in particular the parts which rely
on inclusive trigger selections.
Given the large increase in instantaneous luminosity and trigger efficiency foreseen in \runthree,
this evolution completes a crucial step in allowing the continuation of today's
physics measurements into the future.

The updated Turbo model has already provided a \SI{50}{\percent} reduction in bandwidth in
comparison with saving the full reconstruction, and this saving has been
exploited with the addition of new high-rate trigger selections.
Even larger gains should be possible when applying similar techniques to the
remaining set of trigger selections, and studies are ongoing into these avenues.

\begin{acronym}
  \acro{LHC}{Large Hadron Collider}
  \acro{PV}{primary vertex}
  \acroplural{PV}[PVs]{primary vertices}
  \acro{IP}{impact parameter}
  \acro{EFF}{Event Filter Farm}
\end{acronym}
 
\addcontentsline{toc}{section}{References}
\setboolean{inbibliography}{true}
\bibliographystyle{LHCb}
\bibliography{main,LHCb-PAPER,LHCb-CONF,LHCb-DP,LHCb-TDR}

\newpage

\ifthenelse{\boolean{paperconf}}{
 
\newpage

\centerline{\large\normalfont\bfseries LHCb collaboration}
\begin{flushleft}
\small
A.~N.~Other$^{1}$.\bigskip\newline{\it
\footnotesize
$ ^{1}$University of nowhere\\
}
\end{flushleft}
 
}{}

\end{document}